\documentclass[journal]{IEEEtai}

\usepackage[colorlinks,urlcolor=blue,linkcolor=blue,citecolor=blue]{hyperref}

\usepackage{color,array}

\usepackage{graphicx}
\usepackage{pifont}
\usepackage{siunitx}
\usepackage{booktabs}
\usepackage{algorithm}
\usepackage{cite}
\usepackage{algorithmicx}
\usepackage{algpseudocode}
\usepackage{amsmath,amssymb}
\usepackage{adjustbox}

\usepackage{xspace}
\newcommand{\nprivate}{\textsc{PrivateEdit}\xspace}

\newcommand{\blue}[1]{#1}

\begin{document}

% \title{\nprivate: A Privacy-Preserving Pipeline for Generative Image Editing} 

\title{\nprivate: A Privacy-Preserving Pipeline for Face-Centric Generative Image Editing}

% \if 0
 \author{First A. Author, \IEEEmembership{Fellow, IEEE}, Second B. Author, and Third C. Author, Jr., \IEEEmembership{Member, IEEE}
 \thanks{This paragraph of the first footnote will contain the date on which you submitted your paper for review. It will also contain support information, including sponsor and financial support acknowledgment. For example, ``This work was supported in part by the U.S. Department of Commerce under Grant BS123456.'' }
 \thanks{The next few paragraphs should contain the authors' current affiliations, including current address and e-mail. For example, F. A. Author is with the National Institute of Standards and Technology, Boulder, CO 80305 USA (e-mail: author@boulder.nist.gov).}
 \thanks{S. B. Author, Jr., was with Rice University, Houston, TX 77005 USA. He is now with the Department of Physics, Colorado State University, Fort Collins, CO 80523 USA (e-mail: author@lamar.colostate.edu).}
 \thanks{T. C. Author is with the Electrical Engineering Department, University of Colorado, Boulder, CO 80309 USA, on leave from the National Research Institute for Metals, Tsukuba, Japan (e-mail: author@nrim.go.jp).}
 \thanks{This paragraph will include the Associate Editor who handled your paper.}}
% \fi

\author{
    Dipesh Tamboli,
    Vineet Punyamoorty,
    Atharv Pawar, and
    Vaneet Aggarwal

    \thanks{Dipesh Tamboli was with the Elmore Family School of Electrical and Computer Engineering, Purdue University,  West Lafayette, 47907, Indiana, USA (e-mail: dtamboli@purdue.edu).}
    \thanks{Vineet Punyamoorty is with the Elmore Family School of Electrical and Computer Engineering, Purdue University,  West Lafayette, 47907, Indiana, USA (e-mail: vpunyamo@purdue.edu).}
    \thanks{Atharv Pawar was with the Electrical and Computer Engineering, University of Michigan,  Ann Arbor, 48109, Michigan, USA (e-mail: atharvp@umich.edu).}
    \thanks{Vaneet Aggarwal is with the Edwardson School of Industrial Engineering and the Department of Computer Science, Purdue University, West Lafayette, 47907, Indiana, USA (e-mail: vaneet@purdue.edu).}
    \thanks{This is the author version of the paper accepted in IEEE Transactions on Artificial Intelligence, Feb 2026.}
}

\markboth{IEEE Transactions on Artificial Intelligence,  2026}
{Tamboli \MakeLowercase{\textit{et al.}}: \nprivate: A Privacy-Preserving Pipeline for Generative Image Editing}

\maketitle

\begin{abstract}
Recent advances in generative image editing have enabled transformative applications, from professional head shot generation to avatar stylization. However, these systems often require uploading high-fidelity facial images to third-party models, raising concerns around biometric privacy, data misuse, and user consent. We propose a privacy-preserving pipeline that supports high-quality editing while keeping users in control over their biometric data in face-centric use cases.
Our approach separates identity-sensitive regions from editable image context using on-device segmentation and masking, enabling secure, user-controlled editing without modifying third-party generative models. Unlike traditional cloud-based tools, \nprivate{} enforces privacy by default: biometric data is never exposed or transmitted. This design requires no access to or retraining of third-party models, making it compatible with a wide range of commercial APIs. By treating privacy as a core design constraint, our system supports responsible generative AI centered on user autonomy and trust.
The pipeline includes a tunable masking mechanism that lets users control how much facial information is concealed, allowing them to balance privacy and output fidelity based on trust level or use case. We demonstrate its applicability in professional and creative workflows and provide a user interface for selective anonymization. By advocating privacy-by-design in generative AI, our work offers both technical feasibility and normative guidance for protecting digital identity. The source code is available at \url{https://github.com/Dipeshtamboli/PrivateEdit-Privacy-Preserving-GenAI}.
\end{abstract}

\begin{IEEEImpStatement}
Generative image editing systems have become integral to professional and creative workflows, yet their reliance on cloud-based APIs introduces serious risks to user privacy, particularly when facial data is involved. This work demonstrates that privacy need not come at the expense of utility. By introducing \nprivate, a lightweight, on-device masking pipeline that separates biometric identity from editable context, we show that users can securely leverage third-party generative models without revealing their personal likeness. The proposed approach is model-agnostic, requiring no retraining or internal access, and is thus immediately deployable across existing commercial platforms. Beyond its technical contributions, this work advances the broader agenda of responsible and privacy-by-design generative AI, empowering users with autonomy and transparency in digital identity management.
\end{IEEEImpStatement}


\begin{figure*}[htbp]
\centering
\includegraphics[width=.9\linewidth]{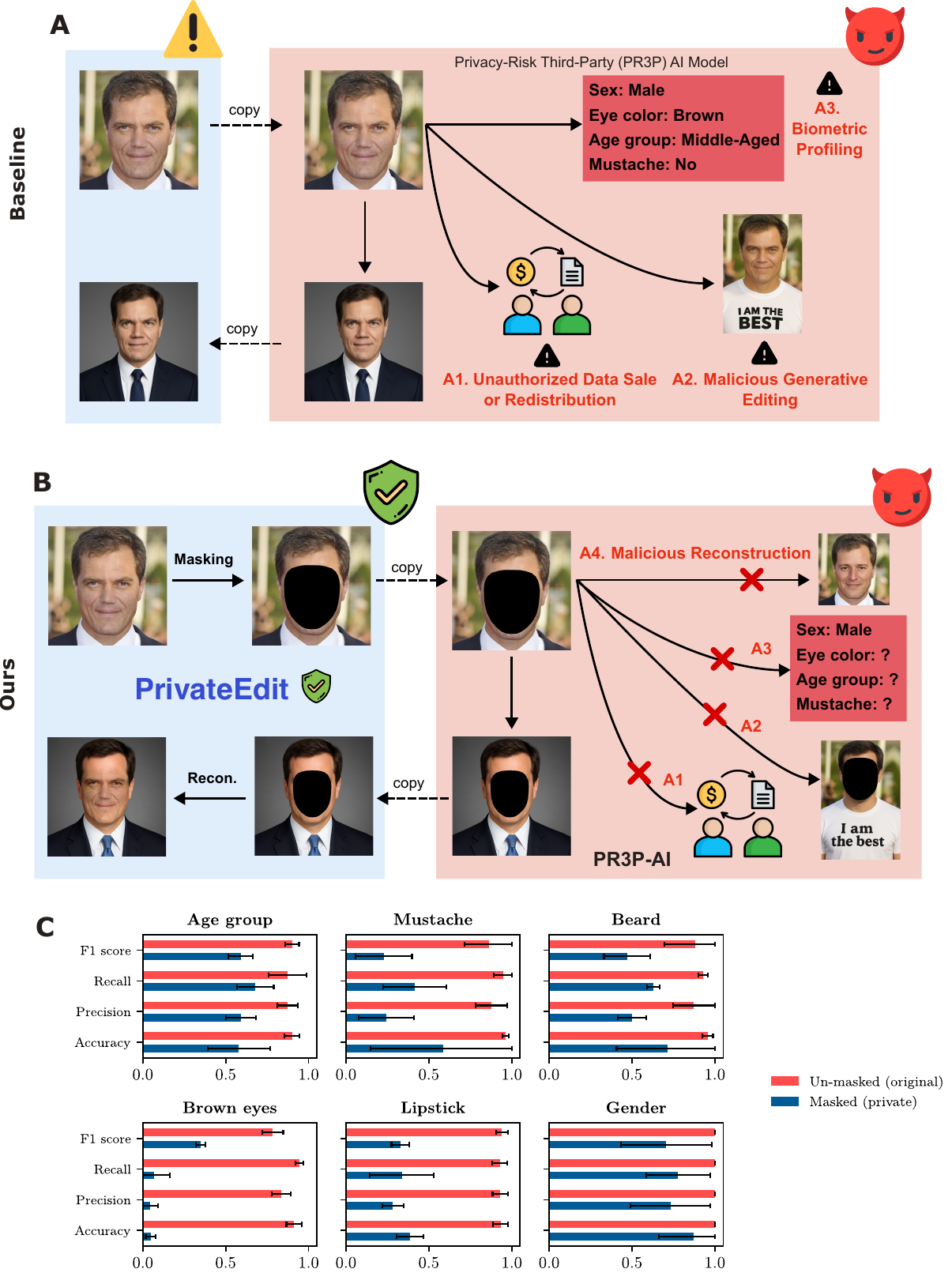}
\caption{\textbf{Quantification of privacy gains.} \textbf{(A)} Overview of our pipeline; \textbf{(B)} Conceptual depiction of privacy threat and masking; \textbf{(C)} Attribute classification performance on masked versus unmasked images, using Gemini, Grok, and LLaMA. Metrics include accuracy, precision, recall, and F1-score. Error bars denote standard deviation across models.}
\label{fig:main-fig}
\end{figure*}

\section{Introduction}

\IEEEPARstart{T}{he} widespread availability of generative artificial intelligence (AI) has enabled transformative applications in image editing, from virtual try-ons \cite{santesteban2021self,wang2022flow,rohil2024fast} to photorealistic portrait synthesis \cite{saharia2022photorealistic,chen2023pixart,bhunia2023person,nichol2021glide}. These capabilities are powered by models such as generative adversarial networks \cite{goodfellow2020generative, goodfellow2014generative} and diffusion models \cite{songscore,ho2020denoising,gaur2025sample,srikanth2025discrete}, which produce high-fidelity results with minimal user input. However, access to such models is increasingly mediated through cloud-based services, as most state-of-the-art systems are proprietary, require large-scale computational resources, depend on fine-tuning with massive internal datasets, or are embedded within commercial APIs that offer no local alternatives~\cite{goodfellow2014generative, rombach2022high, saharia2022photorealistic}.\\
This deployment model poses fundamental risks to user privacy, particularly in applications involving facial data, where uploading images to external servers can expose biometric identifiers to storage, logging, or repurposing beyond user consent~\cite{oleszkiewicz2018sgap, jeong2021ficgan, wang2024edit}. Previous approaches to facial de-identification typically rely on downstream editing or stylization, assume trust in the service provider, or remain vulnerable to reconstruction and attribute inference attacks~\cite{liu2021machine, xiang2024fairness}.
% {Here we propose} a privacy-preserving image editing framework, called \nprivate, that enables users to apply third-party generative models without exposing their biometric identity. Our method employs a lightweight neural network on the user's device to automatically detect and mask the identity-revealing inner facial region before transmitting the image to an external model. After editing, the private region is seamlessly restored using the original data, preserving both output quality and user privacy without modifying or retraining the generative model itself.
% 
% To quantify the privacy gains offered by our pipeline, we introduce a new evaluation protocol that measures the ability of foundation models, including Gemini \cite{team2024gemini}, Grok \cite{xai}, and LLaMA \cite{touvron2023llama}, to infer personal attributes from masked versus unmasked images. We find that classification accuracy and F1 scores for key biometric features drop significantly, often by more than 50\%, on masked inputs, demonstrating that our method meaningfully limits identity inference. By offering a plug-in privacy layer for generative editing, \nprivate establishes a modular, deployable path toward responsible AI, enabling the benefits of powerful generative tools without compromising user autonomy or data sovereignty.

We propose a privacy-preserving image editing framework, called \nprivate{}, that enables users to harness third-party generative models without exposing their biometric identity. Our method uses a lightweight neural network that runs entirely on the user's device to detect and mask identity-sensitive facial regions before transmitting any data to an external model. After editing, the masked region is seamlessly restored using the original image, preserving both output fidelity and user privacy. Crucially, this pipeline requires no access to the internals of the generative model, making it compatible with proprietary, closed-box APIs.

To quantify the privacy gains provided by \nprivate{}, we introduce a model-based evaluation protocol that tests the ability of foundation models, such as Gemini \cite{team2024gemini}, Grok \cite{xai}, and LLaMA \cite{touvron2023llama}, to infer biometric attributes from masked versus unmasked images. We observe a significant drop in inference accuracy and F1 scores across key attributes, often exceeding 50\%, demonstrating that our approach meaningfully limits identity leakage.
Rather than treating privacy as an optional safeguard, \nprivate{} makes it a default, integral part of the editing pipeline. By offering a modular, deployment-compatible privacy layer that works with any generative API without requiring retraining or internal access, \nprivate{} enables responsible, autonomy-preserving use of generative AI, putting users in control of how and when their identity is exposed.

\blue{\nprivate{} focuses on preserving privacy of facial biometric attributes, and does not apply to the protection of non-facial sensitive content in the image. This may include any background objects such as visible documents and text, location-identifying visual landmarks, or non-facial quasi-biometric features such as tattoos or birthmarks visible on exposed body parts.}

\section{Rethinking Privacy in Generative Image Editing Pipelines}\label{sec:intro}

Despite their rapid progress, image-based generative models pose unresolved challenges at the intersection of personalization, automation, and privacy \cite{liu2025generative,golda2024privacy}. Consumer access to generative models is overwhelmingly mediated through commercial APIs, often bundled into applications that edit or synthesize human images, including faces, at scale. These pipelines implicitly assume that high-fidelity edits can only be achieved if the complete, unaltered input image is provided to the generative model. However, this assumption fundamentally limits privacy by design: it demands full disclosure of biometric content to external systems that are neither verifiable nor controllable by the user.

From a privacy engineering perspective, this architecture is problematic. It leaves users exposed not only to direct threats such as private information storage, logging, or data scraping, but also to more subtle forms of inference and modeling leakage \cite{golda2024privacy}. For instance, third-party models may extract latent representations of faces and link them across sessions, or infer demographic traits from images even when it is not part of the editing task. While much attention has been paid to adversarial training and stylization as downstream privacy mitigations \cite{liu2025generative}, these methods are vulnerable to inversion attacks and style leakage. Moreover, they are ill-suited to modern diffusion-based APIs that do not permit internal modification or re-training by the user.

One might ask whether such reliance on external models could be eliminated entirely by running generative models locally on the user's device. While appealing in theory, this approach fails in practice for several reasons. First, many state-of-the-art diffusion models are proprietary and not publicly available in full precision or trained form. Second, even open-source models typically require GPU memory and compute budgets far beyond what consumer-grade devices can provide. For example, high-quality inpainting or style transfer using a diffusion model may require multiple gigabytes of VRAM, which is prohibitive for common smartphones. Third, hosting local models shifts maintenance and storage burdens to the user, complicates updates, and decouples users from curated improvements or safety interventions made by developers. As a result, the dominant design pattern has become one in which generative intelligence is delivered through remote APIs, thus cementing the privacy threat.

Other potential mitigations, such as post-edit de-identification or feature blurring, also fall short. Post-edit masking is lossy and irreversible once biometric data has been uploaded; blurring-based techniques can preserve enough spatial structure for attribute inference, as shown in prior work~\cite{wang2024edit}. Stylization-based anonymization may distort facial characteristics but can often be reversed or sidestepped by adversarial reconstruction attacks. In short, most existing methods either rely on full trust in the editing service, degrade image quality, or require deep integration with the generative model itself, making them incompatible with the prevailing cloud-based API paradigm.

Critically, many such solutions assume a cooperative or at least transparent generative model. But this assumption breaks down when the model is a black-box service with unknown internal logging, fine-tuning, or retention policies. This misalignment between the privacy threat model and real-world deployment constraints has created a practical gap: users must choose between full image editing functionality and full biometric privacy. As generative APIs become increasingly commodified, this tradeoff is no longer tenable.

Our work addresses this problem by designing an upstream privacy mechanism that is model-agnostic, deployment-compatible, and quantifiably effective. We introduce a lightweight neural network that runs entirely on-device and detects the minimal inner facial region required to conceal biometric identity. This region is masked—typically via simple occlusion—before any part of the image is transmitted. Crucially, the masking network is optimized to run efficiently on compute-poor edge devices such as mobile phones, requiring neither user supervision nor internet access. The masked image is then sent to the generative API, which performs the requested edit (e.g., professional headshot rendering), after which the masked region is seamlessly restored using the original image. In this pipeline, the untrusted generative model never sees the identity-revealing data, and the user retains full control over the biometric core of the image.

To validate the privacy benefit of this design, we formulate a model-based privacy inference test. Unlike prior methods that rely on structural similarity or identity verification scores \cite{jeong2021ficgan, oleszkiewicz2019siamese}, our evaluation quantifies the ability of state-of-the-art foundation models to infer a fixed set of biometric attributes from masked versus unmasked images. This includes categorical attributes such as gender, presence of facial hair, eye color, and age group. By aggregating predictions from models including Gemini, Grok, and LLaMA, we are able to simulate realistic inference attacks under plausible adversarial conditions. Notably, the classification accuracy, precision, and F1 score for multiple attributes drop substantially, often by more than half, when the masked pipeline is used. These results suggest that upstream masking meaningfully reduces biometric leakage even under strong adversarial assumptions. While these results illustrate significant gains in face-based editing, our method currently applies only to portrait-style image pipelines and is not designed to generalize to non-facial biometric or arbitrary visual data. 

This section motivates and grounds our method not as a substitute for secure computation or differential privacy, but as a practical, modular component of the broader privacy toolkit. Our goal is to enable privacy-aligned generative experiences without requiring trust in the model, retraining of its internals, or access to proprietary infrastructure. The rest of the paper builds on this foundation by describing the system architecture, experimental methodology, quantitative results, and deployment implications.

\blue{\section{
Privacy Threat Model and Information-Theoretic Analysis
}}
% \subsection{Privacy Threat Model and Information-Theoretic Analysis}
\label{sec:privacy_theory}

\blue{
\paragraph{Threat Model}
We consider a cloud-based image editing setting with an \emph{honest-but-curious} adversary. The adversary has full access to the image transmitted to the cloud and the text prompt used for editing, and is aware of the overall pipeline design. However, the adversary does \emph{not} have access to the original image, facial masks, facial landmarks, or the local reintegration module executed on the user device. The adversary’s objective is to infer the biometric identity of the individual appearing in the input image.
}

\blue{
\paragraph{Explicit Facial Removal}
In the \nprivate pipeline, the facial region is explicitly removed prior to cloud transmission. Let \( I \in \mathbb{R}^{H \times W \times 3} \) denote the original image, and let \( M \in \{0,1\}^{H \times W} \) be a binary facial mask obtained via face detection and landmark localization, where \( M=1 \) denotes facial pixels. The masked image transmitted to the cloud is defined as
\begin{equation}
I_m = (1 - M) \odot I + M \odot c,
\end{equation}
where \( \odot \) denotes element-wise multiplication and \( c \) is a constant mask value (e.g., zero-valued pixels). This operation deterministically removes all facial pixel information from the image prior to cloud inference.
}

\blue{
\paragraph{Biometric Privacy Definition}
Let \( z_{\text{id}} = f_{\text{FR}}(I) \) denote an identity-specific biometric embedding extracted by any face recognition model \( f_{\text{FR}} \). We define biometric privacy as the inability of any adversary to infer \( z_{\text{id}} \) or the corresponding identity label from the transmitted image \( I_m \). Under the assumption that biometric identity information is localized within the facial region, explicit facial removal yields
\begin{equation}
\mathbb{I}(z_{\text{id}}; I_m) = 0,
\end{equation}
where \( \mathbb{I}(\cdot;\cdot) \) denotes mutual information. Consequently, the success probability of any identity inference attack from the cloud-visible image is upper-bounded by chance-level guessing based on non-facial context alone.
}

\blue{
\paragraph{Soft Attribute Leakage}
While facial identity is fully removed, some non-identifying semantic attributes may still be inferred from non-facial context. We decompose semantic attributes as
\begin{equation}
z_{\text{attr}} = (z_{\text{face-attr}}, z_{\text{context-attr}}),
\end{equation}
where \( z_{\text{face-attr}} \) includes facial attributes such as eye color or facial hair, and \( z_{\text{context-attr}} \) includes attributes inferable from hair, clothing, or body shape. Since all facial pixels are removed,
\begin{equation}
\mathbb{I}(z_{\text{face-attr}}; I_m) = 0,
\end{equation}
while
\begin{equation}
\mathbb{I}(z_{\text{context-attr}}; I_m) > 0.
\end{equation}
This formulation explains the empirical results in Fig.~\ref{fig:main-fig}, where facial attributes collapse to near-random performance under masking, while coarse attributes such as gender remain partially inferable due to contextual cues.
}

\blue{
\paragraph{Cloud Editing and Local Reintegration}
The masked image \( I_m \) is processed by a cloud-based generative editor \( \mathcal{G} \) together with a text prompt \( \mathcal{P} \), producing an edited image
\begin{equation}
I_e = \mathcal{G}(I_m, \mathcal{P}).
\end{equation}
The original facial content is reintroduced locally using a reintegration operator \( \mathcal{R} \),
\begin{equation}
I_f = \mathcal{R}(I_e, I, M),
\end{equation}
which is executed entirely on the user device. Neither the original image \( I \) nor the mask \( M \) is ever transmitted to the cloud, ensuring that no facial biometric information is exposed during cloud-based processing.
}

\blue{
\paragraph{Scope and Limitations}
The privacy guarantees of \nprivate{} hold under accurate facial mask coverage and are limited to facial biometric identity. Attributes encoded outside the face region or unique contextual identifiers fall outside the scope of facial privacy protection and can be potential sources of background information leakage.
}
\blue{
\section{
Geometric Validity and Safe Editing Bounds
}}
\label{sec:geometry_bounds}

\blue{
\nprivate{} relies on local facial reintegration, which assumes geometric consistency between the original image and the cloud-edited output. This assumption is intentional and aligned with the target application of professional portrait and headshot editing, where both input images and desired outputs are expected to be approximately front-facing.
}

\blue{
\paragraph{Geometric Assumptions}
Let \( I \) denote the original image and \( I_e \) the edited image returned by the cloud-based generative model. We assume that the facial region in \( I_e \) preserves the 2D facial topology of \( I \) up to an affine transformation. This corresponds to bounded variations in head pose, expression, and camera orientation. Formally, let
\[
\Delta \boldsymbol{\theta} = (\Delta \text{yaw}, \Delta \text{pitch}, \Delta \text{roll})
\]
denote the change in head orientation induced by the editing process. Reintegration remains well-defined when
\[
\lVert \Delta \boldsymbol{\theta} \rVert \le \tau,
\]
where \( \tau \) is an application-dependent tolerance within which facial correspondences remain approximately affine in the image plane.
}

\blue{
\paragraph{Failure Modes}
\nprivate{} is designed for privacy-preserving editing under \emph{controlled geometric conditions} and does not aim to support camera viewpoint or head-pose changes. Edits that induce large out-of-plane rotations, non-frontal camera angles, significant viewpoint or head-pose changes violate the geometric assumptions. In such cases, standard 2D geometric alignment and inverse warping cannot recover the underlying 3D facial structure, and reintegration quality may degrade. \nprivate{} does not claim robustness to arbitrary 3D pose transformations, and such scenarios are explicitly considered outside the intended operating regime.
}

\blue{
\paragraph{Ensuring Application Alignment}
From a user utility perspective, these constraints are consistent with real-world professional headshot workflows, where subjects are typically captured in near-frontal poses. If the input image does not satisfy this assumption, users can be guided to capture an image within the supported pose range. Additionally, if the generative model produces an output with an unintended orientation, \nprivate{} can issue a follow-up prompt requesting a front-facing view, which is supported by modern generative editing models. These design choices ensure that geometric consistency is maintained without exposing facial information to the cloud.
}

%\blue{
%\paragraph{Scope and Limitations}
%Potential avenues for investigation includes requesting multiple images of the user in different head-poses and using 3-D reconstruction .
%By explicitly defining the safe editing bound, we delineate the validity scope of the reintegration mechanism and clarify the conditions under which the method is theoretically and empirically sound.
%}

\section{Evaluation Setup and Baselines}\label{sec2}

\begin{table}[ht]
\centering
\setlength{\tabcolsep}{1mm} % reduce column spacing
\caption{Quantitative comparison of Face‐FID, Cosine similarity, and CLIP alignment. Lower Face‐FID and higher Cosine indicate better facial similarity; higher CLIP reflects better alignment with the “studio headshot” prompt. “Privacy” indicates biometric preservation (\ding{51}), not preserved (\ding{55}), or NA.}
\label{tab:metrics_2dec}
\begin{tabular}{l 
                S[table-format=3.2] 
                S[table-format=1.2] 
                S[table-format=1.2] 
                c}
\toprule
\textbf{Method} & \textbf{Face‐FID} & \textbf{Cosine} & \textbf{CLIP} & \textbf{Privacy} \\
\midrule
Input                  & 0.00 & 1.00 & 0.25 & NA \\
GPT (No Privacy)       & 517.05 & 0.39 & 0.30 & \ding{55} \\
GPT Reconstructed      & 683.04 & 0.12 & 0.30 & \ding{51} \\
\textbf{Ours}          & 226.50 & 0.77 & 0.29 & \ding{51} \\
\bottomrule
\end{tabular}
\end{table}

\begin{table}[ht]
\centering
\setlength{\tabcolsep}{1mm} % reduce column spacing
\caption{ 
Ablation study on the effect of mask ratio in \nprivate{}. Face-FID, Cosine similarity, and CLIP score measure visual fidelity and task alignment (higher cosine / clip and lower Face-FID are better). Privacy is quantified via attribute inference F1-scores (lower is better). Decreasing the mask ratio improves editability but increases biometric leakage.}
\label{tab:mask_ratio_full_ablation}
\resizebox{\linewidth}{!}{
\begin{tabular}{l 
                S[table-format=3.2]
                S[table-format=1.2]
                S[table-format=1.2]
                S[table-format=1.2]
                S[table-format=1.2]
                S[table-format=1.2]
                S[table-format=1.2]
                S[table-format=1.2]}
\toprule
\textbf{Mask Ratio} &
\textbf{Face-FID} &
\textbf{Cosine} &
\textbf{CLIP} &
\textbf{Age} &
\textbf{Mustache} &
\textbf{Beard} &
\textbf{Brown Eyes} &
\textbf{Gender} \\
\midrule
0.60 & 198.40 & 0.83 & 0.31 & 0.71 & 0.62 & 0.66 & 0.58 & 0.89 \\
0.75 & 212.10 & 0.81 & 0.30 & 0.64 & 0.51 & 0.57 & 0.44 & 0.87 \\
0.90 & 221.30 & 0.78 & 0.29 & 0.56 & 0.41 & 0.48 & 0.29 & 0.85 \\
\textbf{1.00} & \textbf{226.50} & \textbf{0.77} & \textbf{0.29} & \textbf{0.52} & \textbf{0.36} & \textbf{0.50} & \textbf{0.21} & \textbf{0.84} \\
\bottomrule
\end{tabular}
}
\end{table}

\begin{table}[htbp]
\centering
\blue{
\caption{Mean Opinion Scores (MOS) from User Study ($N=10$). Scores on a 5-point Likert scale (mean $\pm$ std. dev). Higher is better.}
\label{tab:user_study}
\resizebox{\columnwidth}{!}{%
\setlength{\tabcolsep}{1.2mm}
\begin{tabular}{@{}lcc@{}}
\toprule
\textbf{Evaluation Criteria} & \textbf{Baseline (Cloud API)} & \textbf{PrivateEdit (Ours)} \\ \midrule
Privacy (Pre-transmission)\textsuperscript{*} & $1.05 \pm 0.21$ & $\mathbf{4.95 \pm 0.08}$ \\
Identity Preservation & $4.72 \pm 0.31$ & $\mathbf{4.96 \pm 0.38}$ \\
Blending Naturalness & $\mathbf{4.88 \pm 0.19}$ & $4.72 \pm 0.22$ \\ \bottomrule
\addlinespace
\multicolumn{3}{l}{\small \textit{*Privacy score: 5 = Unidentifiable (Best), 1 = Identifiable (Worst).}} \\
\end{tabular}%
}
}
\end{table}

% \begin{table}[ht]
% \centering
% \caption{Quantitative comparison of Face‐FID, Cosine similarity, and CLIP alignment. Lower Face‐FID and higher Cosine indicate better facial similarity; higher CLIP reflects better alignment with the “studio headshot” prompt. “Privacy” indicates biometric preservation (\ding{51}), not preserved (\ding{55}), or NA.}
% \label{tab:metrics_2dec}
% \resizebox{\columnwidth}{!}{
% \begin{tabular}{l 
%                 S[table-format=3.2] 
%                 S[table-format=1.2] 
%                 S[table-format=1.2] 
%                 c}
% \toprule
% \textbf{Method} & \textbf{Face‐FID} & \textbf{Cosine} & \textbf{CLIP} & \textbf{Privacy} \\
% \midrule
% Input                  & 0.00 & 1.00 & 0.25 & NA \\
% GPT (No Privacy)       & 517.05 & 0.39 & 0.30 & \ding{55} \\
% GPT Reconstructed      & 683.04 & 0.12 & 0.30 & \ding{51} \\
% \textbf{Ours}          & 226.50 & 0.77 & 0.29 & \ding{51} \\
% \bottomrule
% \end{tabular}
% }
% \end{table}

% \section{Results}

We evaluate the effectiveness of our privacy-preserving image editing pipeline using the CelebA dataset \cite{liu2015deep}, a benchmark widely used for facial attribute manipulation. Our experiments are designed to assess both the visual quality and privacy protection offered by the proposed method. To do so, we compare three editing configurations, each representing a distinct trade-off between utility and privacy:

\begin{itemize}
    \item \texttt{GPT (No Privacy)}: The original, unmasked image is sent directly to the ChatGPT Vision API \cite{chatgptvisionapi} with the prompt \textit{“Make a professional headshot out of this picture”}. This setting provides the best-case performance for utility, but offers no privacy protection.
    \item \texttt{GPT (Reconstructed)}: A masked version of the image is sent to the same API. This simulates an adversarial reconstruction attempt by the third-party model, which receives no identifying facial features. The output represents what the model is able to reconstruct from the occluded input alone.
    \item \nprivate{} \texttt{(Ours Reintegrated)}: Our full \nprivate{} pipeline. The masked image is first sent to the generative model, and the result is then post-processed by locally reinserting the original, private facial region. This configuration preserves privacy during editing while restoring identity after the generative step.
\end{itemize}

\section{Visual Fidelity and Identity Preservation}

We evaluate each of the three editing configurations using a set of complementary metrics, each designed to capture a distinct aspect of output quality and identity preservation:

\begin{itemize}
    \item \textbf{FID-Face}: This is a variant of the Fréchet Inception Distance \cite{heusel2017gans}, computed specifically on facial feature embeddings rather than whole-image features. It evaluates the perceptual quality of the generated face by comparing the statistical distribution of embeddings between original and edited images. Lower FID-Face scores indicate higher visual realism and better coherence in the facial region, making it well-suited for detecting subtle differences in facial structure.

    \item \textbf{Cosine similarity}: This metric compares the facial embeddings of the input and output images to assess identity retention. It quantifies how similar the representations of the face remain after editing. A high cosine similarity implies that the key identity characteristics are preserved, which is critical when the goal is to maintain a faithful depiction of the subject.

    \item \textbf{CLIP score}: This metric leverages a pre-trained vision-language model, CLIP \cite{radford_learning_2021}, to evaluate semantic alignment between the edited image and the intended editing instruction. It computes the similarity between the output image’s embedding and the embedding of the textual prompt. A higher CLIP score indicates that the output successfully reflects the desired transformation as described by the prompt.
\end{itemize}

\begin{figure*}[htbp]
\centering
\includegraphics[width=0.75\linewidth]{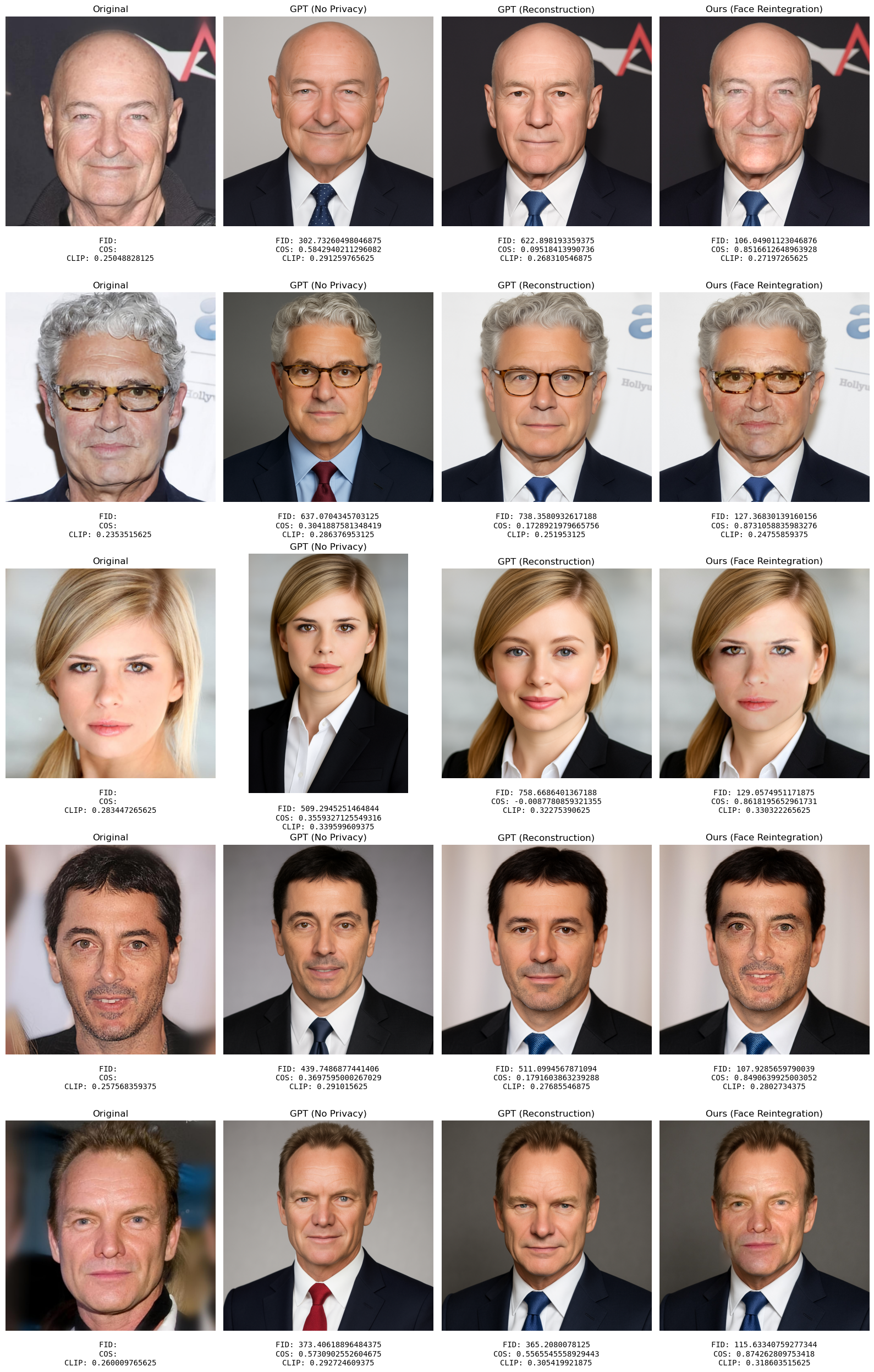}
\caption{\textbf{Qualitative comparison of privacy‐preserving face edits.} Each row corresponds to a different input image (leftmost column), followed by three edited versions: “GPT (No Privacy)” (second column), “GPT (Reconstruction)” (third column), and our method (“Ours (Face Re‐integration)”, rightmost column). Beneath each image, we report Face‐FID (FID), Cosine similarity (COS), and CLIP score (CLIP) relative to the prompt “Convert this image into a professional studio headshot.”}
\label{fig:qualitative_results}
\end{figure*}

Figure~\ref{fig:qualitative_results} presents a visual comparison of outputs from the three configurations, along with metric values for each example. Quantitative results, averaged across the dataset, are summarized in Table~\ref{tab:metrics_2dec}. Among all methods, our approach achieves the lowest FID-Face score (226.5), indicating the most realistic and coherent facial edits. This reflects better visual quality and preservation of facial features, which are important for applications like profile photo generation or professional headshots. The \texttt{GPT (No Privacy)} baseline results in a significantly higher FID of 517.1, suggesting that even unmasked input does not always lead to photorealistic or identity-consistent edits. The adversarial \texttt{GPT (Reconstructed)} setting performs worst with a score of 683.0, underscoring that identity-masked inputs are not easily reconstructible by third-party models.

In terms of identity preservation, our method yields the highest cosine similarity (0.766), substantially outperforming both the unmasked baseline (0.389) and the adversarial reconstruction attempt (0.123). This suggests that our reintegration step effectively restores user identity in the final output, an essential property for applications where personalized representation is important, such as headshot generation.

Finally, our CLIP score (0.289) is comparable to that of \texttt{GPT (No Privacy)} (0.304), confirming that our privacy-preserving intervention does not disrupt the model’s ability to follow the editing prompt. This alignment ensures that the intended style and semantic attributes are retained, making the edited image not just private but also functionally useful.

\blue{
\paragraph{Effect of Mask Ratio on Privacy-Utility Trade-off}
\nprivate{} enforces facial privacy by explicitly removing facial pixels prior to cloud-based inference. Consequently, the primary controllable factor governing the trade-off between editability and privacy is the extent of facial masking. To analyze this effect, we conduct an ablation study varying the facial mask ratio, as summarized in Table~\ref{tab:mask_ratio_full_ablation}. Lower mask ratios preserve more facial context, enabling the generative model to better capture scene semantics and lighting, which leads to improved visual fidelity and prompt alignment, as reflected by lower Face-FID and higher cosine similarity and CLIP scores. However, this comes at the cost of increased biometric attribute leakage, evidenced by higher attribute inference F1-scores across age, facial hair, eye color, and gender. Conversely, full facial masking maximizes biometric privacy by suppressing attribute inference while incurring only a modest degradation in editing quality. These results demonstrate that mask ratio serves as an explicit and interpretable control knob for navigating the privacy-utility balance in \nprivate{}.
}
\\

\blue{ \paragraph{Human Evaluation} A significant finding of our evaluation is that \nprivate{} outperforms the baseline Cloud API in identity preservation. As seen in Table \ref{tab:metrics_2dec}, the GPT (No Privacy) baseline achieves a Cosine Similarity of only 0.39. This is due to 'identity drift,' where the generative model modifies the user's facial structure to align with the semantic prompt (e.g., making the face more symmetrical or 'professional'). Because \nprivate{} performs local reintegration of the original biometric pixels (Equation 7), it bypasses this generative drift entirely. The user study results (Table \ref{tab:user_study}) verify that human raters find \nprivate{}'s outputs significantly more representative of the original subject than the baseline. }

\blue{ Regarding visual quality, the baseline Cloud API slightly outperformed our method in \textit{Blending Naturalness}. This is expected, as the baseline utilizes global neural synthesis to generate the face and background simultaneously. In contrast, while our Poisson blending effectively handles ambient color shifts, a small subset of \nprivate{} results exhibited minor lighting inconsistencies, particularly in cases with strong directional shadows that contradicted the original face's illumination. However, participants noted that these artifacts were often negligible for casual use cases. Overall, the study highlights a dual advantage: \nprivate{} ensures absolute biometric privacy while simultaneously providing higher identity fidelity, offering a superior balance between security and utility than unconstrained generative models. }

% \begin{table}[ht]
% \centering
% \caption{Evaluation of editing outputs across fidelity, identity, and prompt alignment metrics. Lower FID-Face and higher cosine/CLIP scores indicate better performance.}
% \label{tab:metrics_2dec}
% \begin{tabular}{lccc}
% \toprule
% Method & FID-Face ↓ & Cosine Similarity ↑ & CLIP Score ↑ \\
% \midrule
% Input (Reference) & 0.00 & 1.000 & 0.250 \\
% GPT (No Privacy) & 517.05 & 0.389 & 0.304 \\
% GPT (Reconstructed) & 683.04 & 0.123 & 0.262 \\
% Ours (Reintegrated) & \textbf{226.50} & \textbf{0.766} & 0.289 \\
% \bottomrule
% \end{tabular}
% \end{table}

% \begin{table}[ht]
% \centering
% \setlength{\tabcolsep}{1mm} % reduced column spacing
% \caption{Evaluation of editing outputs across fidelity, identity, and prompt alignment metrics. Lower FID and higher cosine/CLIP scores indicate better performance.}
% \label{tab:metrics_2dec}
% \begin{tabular}{lccc}
% \toprule
% \textbf{Method} & \textbf{FID-Face} ↓ & \textbf{Cosine Sim.} ↑ & \textbf{CLIP} ↑ \\
% \midrule
% Input (Reference)      & 0.00    & 1.000 & 0.250 \\
% GPT (No Privacy)       & 517.05  & 0.389 & \textbf{0.304} \\
% GPT (Reconstructed)    & 683.04  & 0.123 & 0.262 \\
% Ours (\nprivate{})      & \textbf{226.50}  & \textbf{0.766} & 0.289 \\
% \bottomrule
% \end{tabular}
% \end{table}

\section{Quantifying Privacy Gains through Attribute Inference}

To assess the privacy impact of our masking strategy, we introduce a biometric inference benchmarking scheme using large foundation models. The objective is to determine how much sensitive attribute information remains inferable from the masked images, thereby quantifying the privacy gains introduced by our pipeline. We prompt three state-of-the-art vision-language models (Gemini~\cite{team2024gemini}, Grok~\cite{xai}, and LLaMA~\cite{touvron2023llama}) to predict a set of seven biometric attributes. These include categorical traits such as age group, gender, eye color, and presence of facial hair, which serve as proxies for identity-revealing information. For each configuration (original vs.\ masked), we measure the classification performance using four standard metrics: accuracy, precision, recall, and F1-score.

We adopt standard attribute-classification metrics that directly measure how well a model can infer biometric traits from masked versus unmasked inputs. These metrics capture different aspects of inference reliability, allowing us to quantify privacy gains as reductions in predictive performance:
\begin{itemize}
    \item \textbf{Accuracy}: Reports the overall proportion of correct predictions, providing a broad indication of how much identifiable information remains accessible after masking.
    \item \textbf{Precision}: Measures the fraction of predicted positives that are correct, reflecting whether the model can still make confident attribute assertions without direct facial evidence.
    \item \textbf{Recall}: Evaluates the proportion of true positives successfully identified, highlighting the extent to which masking disrupts the model’s ability to detect attributes that are actually present.
    \item \textbf{F1-score}: The harmonic mean of precision and recall, offering a balanced summary metric that is especially informative in cases where occlusion alters class imbalance or model confidence.
\end{itemize}

We perform inference using three high-capacity foundation models (Gemini~\cite{team2024gemini}, Grok~\cite{xai}, and LLaMA~\cite{touvron2023llama}) that together span both proprietary and open-source ecosystems and are developed by different organizations. This choice reflects the diversity of real-world deployment scenarios, where external editing or analysis services may rely on either closed commercial APIs or publicly available models. By evaluating privacy leakage across this heterogeneous set of systems, we obtain a broader and more representative picture of how much biometric information can still be extracted from masked inputs in practice.

We summarize the inference results in Figure~\ref{fig:main-fig}C. We observe a substantial overall decline in attribute inference performance when predictions are made on masked images instead of unmasked ones. However, the degree of degradation varies by attribute, revealing important insights into the effectiveness and limitations of our masking approach. For instance, we find that gender classification remains largely unaffected by masking. This suggests that gender can often be inferred from features outside the masked area, such as hairstyle, jawline, or clothing. This represents a limitation of our pipeline, as gender is a relatively high-level semantic feature that may not be fully obfuscated through geometric masking alone.

In contrast, attributes like eye color, lipstick presence, and facial hair exhibit significant privacy gains. The average accuracy for brown eye detection falls from 0.91 to just 0.05 when the input is masked, indicating that the models are no longer able to rely on direct pixel evidence for iris color. Similarly, the F1-score for mustache detection drops from 0.88 to 0.36, and for beard detection from 0.87 to 0.50, suggesting that lower-face obfuscation effectively prevents inference of these attributes. Lipstick prediction also suffers a notable decline in performance, further validating that our masked regions contain the key signal required for these predictions.

We also observe a moderate but consistent reduction in performance for age group classification. The F1-score for this task declines, though not as sharply as for facial hair or eye color. This may be attributed to the fact that models can use indirect cues such as gray hair, skin texture, or head shape, which often remain visible outside the masked region. 

Taken together, these findings illustrate the selective strength of our approach. The \nprivate{} pipeline effectively prevents inference of attributes that are spatially localized to the masked region, while offering partial protection against more distributed traits. Despite this limitation, our strategy achieves a significant reduction in biometric leakage across most key indicators. Moreover, the masking process is model-agnostic and requires no access to or modification of the underlying generative model, making the solution broadly compatible with real-world applications. By allowing high-quality image editing without fully exposing sensitive features, our method balances utility and privacy in a deployable and transparent way.

\section{Methodology}

In this section, we describe the \nprivate{} pipeline for privacy‐preserving generative image editing. The overall workflow is illustrated in Figure~\ref{fig:main-fig}, and summarized in Algorithm~\ref{alg:privacy_editing}. Our goal is to enable arbitrary user‐specified edits (e.g., “convert to studio headshot”) while ensuring that no sensitive identity information leaves the user’s device. The main steps are: (1) local detection and masking of identity‐sensitive facial regions, (2) optional demonstration of reconstruction failure, (3) third‐party editing of the masked image according to a user‐provided task, (4) re‐injection of the original identity into the edited result, and (5) automated evaluation of task‐alignment via CLIP scoring.

\subsection{Data}
The CelebFaces Attributes Dataset (CelebA)~\cite{liu2015faceattributes} consists of images of celebrity individuals annotated with their facial attributes. To construct the test-set used for our experiments, we identified a subset of images where the subject is facing towards the camera, then chose a random sample of 100 images from this subset for evaluation.

\begin{algorithm}[h]
\caption{\nprivate{}: Privacy-Preserving Generative Editing}
\label{alg:privacy_editing}
\begin{adjustbox}{width=.95\linewidth}
\begin{minipage}{\linewidth}
\begin{algorithmic}[1]
\Require
    \Statex $\mathit{img\_id}$: Unique image identifier
    \Statex $\mathit{edit\_task}$: Text prompt (e.g., "studio headshot")
\Ensure
    \Statex Identity-restored edited output and CLIP alignment scores

\Procedure{RunPipeline}{$\mathit{img\_id}, \mathit{edit\_task}$}
    \State $\mathit{inp} \gets$ \texttt{"input\_og\_imgs/"} $+\,\mathit{img\_id} +$ \texttt{".jpg"}
    \State $\mathit{mask} \gets$ \texttt{"masked\_imgs/"} $+\,\mathit{img\_id} +$ \texttt{"\_mask.jpg"}
    \State $\mathit{gen\_private} \gets$ \texttt{"gpt\_generated/"} $+\,\mathit{img\_id} +$ \texttt{"\_private.jpg"}
    \State $\mathit{gen\_recon} \gets$ \texttt{"gpt\_generated/"} $+\,\mathit{img\_id} +$ \texttt{"\_recon.jpg"}
    \State $\mathit{final} \gets$ \texttt{"ours\_edited/"} $+\,\mathit{img\_id} +$ \texttt{"\_final.jpg"}

    \State \textbf{// Face occlusion}
    \State \Call{MaskImage}{$\mathit{inp}, \mathit{mask}$}

    \State \textbf{// Reconstruction via masked input}
    \State \Call{CallThirdPartyAPI}{$\mathit{mask}, \mathit{gen\_recon}, \mathit{edit\_task}$}

    \State \textbf{// Third-party editing}
    \State \Call{CallThirdPartyAPI}{$\mathit{mask}, \mathit{gen\_private}, \mathit{edit\_task}$}

    % \If{\Call{exists}{$\mathit{final}$}} \Return \texttt{"Skip"} \EndIf

    \State \textbf{// Identity reinjection}
    \State \Call{SwapFaceBack}{$\mathit{inp}, \mathit{gen\_private}, \mathit{final}$}

    \State \textbf{// CLIP-based evaluation}
    \State $s_{\text{orig}} \gets \Call{ComputeCLIPScore}{\mathit{inp}, \mathit{edit\_task}}$
    \State $s_{\text{gpt}} \gets \Call{ComputeCLIPScore}{\mathit{gen\_private}, \mathit{edit\_task}}$
    \State $s_{\text{ours}} \gets \Call{ComputeCLIPScore}{\mathit{final}, \mathit{edit\_task}}$

    % \State \Call{PlotImagePipeline}{$\mathit{img\_id}$} \Comment{(Optional)}

    \State \Return $\{s_{\text{orig}}, s_{\text{gpt}}, s_{\text{ours}}\}$
\EndProcedure

% \Procedure{MainLoop}{}
%     \State $\mathit{img\_list} \gets \Call{listFiles}{\texttt{"input\_og\_imgs/"}}$
%     \State $\mathit{edit\_task} \gets$ user-defined prompt
%     \ForAll{$\mathit{img\_id} \in \mathit{img\_list}$}
%         \State $\{s_{\text{orig}}, s_{\text{gpt}}, s_{\text{ours}}\} \gets \Call{RunPipeline}{\mathit{img\_id}, \mathit{edit\_task}}$
%         \State \Call{storeScores}{$\mathit{img\_id}, s_{\text{orig}}, s_{\text{gpt}}, s_{\text{ours}}$}
%     \EndFor
% \EndProcedure
\end{algorithmic}
\end{minipage}
\end{adjustbox}
\end{algorithm}

% \subsection{Algorithm Overview}

% Figure~\ref{fig:main-fig} and Algorithm~\ref{alg:privacy_editing} outline the end‐to‐end workflow of \nprivate{}. Given any input image (identified by \texttt{image\_id}) and a user‐specified editing prompt (the “task”), \nprivate{} ensures that sensitive facial features never leave the user’s device. First, \nprivate{} automatically detects and masks identity‐bearing facial regions using a combination of MediaPipe’s \cite{lugaresi2019mediapipe} landmark extraction, convex‐hull generation, and edge blending (\textbf{Step 2}). Next, an optional reconstruction check (\textbf{Step 3}) demonstrates that a masked face cannot be recovered by a malicious third‐party model. The masked image is then sent to any third‐party generative API along with the user’s desired task (e.g., “convert to studio headshot”) (\textbf{Step 4}). Because the input is occluded, no actual identity information is exposed. Once the third‐party returns an edited, masked output, \nprivate{} locally re‐injects the original face by aligning and blending the unmasked facial region (\textbf{Step 5}). Finally, an automated CLIP‐score evaluation (\textbf{Step 6}) quantifies how well the original, baseline (third‐party), and final \nprivate{} outputs satisfy the user’s prompt. An optional visualization (\textbf{Step 7}) collates all stages into a composite figure for debugging or presentation.

\subsection{Algorithm Overview}

Figure~\ref{fig:main-fig} and Algorithm~\ref{alg:privacy_editing} present the end-to-end pipeline of \nprivate{}. Given an input image (\texttt{img\_id}) and a user-defined edit prompt (\texttt{edit\_task}), the method preserves biometric privacy while enabling third-party image generation.

The pipeline begins with local face occlusion (lines 6–7), where facial landmarks are detected and masked using convex-hull generation and edge blending via MediaPipe~\cite{lugaresi2019mediapipe}. An optional reconstruction vulnerability check (line 9) uses the masked input to assess its resilience to third-party recovery. The core editing step (line 11) invokes the generative API using the masked image and the user prompt.

After receiving the edited output, identity reinjection (line 14) is performed by aligning and blending the original unmasked face into the generated image. CLIP-based alignment evaluation (lines 17–19) then scores the original, third-party, and final outputs with respect to the prompt. Optionally, a composite visualization (line 21) is generated to summarize the pipeline output for inspection or presentation.

\subsection{Face Detection and Masking}

Accurate facial‐landmark detection is critical to preserve privacy. We employ MediaPipe \cite{lugaresi2019mediapipe} FaceMesh to locate approximately 468 3D points on the user’s face. From these landmarks, \nprivate{} constructs a tight convex‐hull around the identity‐sensitive region (cheeks, nose, eyes, and mouth). To ensure robustness, the hull is scaled by a small factor (controlled internally, but not exposed to the end user) to cover any residual identity cues. A Canny‐edge overlay is blended smoothly onto this hull to create a soft occlusion, guaranteeing that downstream generative models never receive an unprotected face. This entire procedure is implemented in a single call: 
\[
  \Call{MaskImage}{\mathit{inpPath},\,\mathit{maskPath}}.
\]
The resulting \texttt{maskPath} image retains all non‐facial content (hair, background, clothing) while fully concealing biometric features.

% \subsection{Demonstration of Reconstruction Robustness (Optional)}

% Although the primary goal is to never send sensitive data out, \nprivate{} optionally illustrates the strength of its mask by calling:
% \[
%   \Call{CallThirdPartyAPI}{\mathit{maskPath},\,\mathit{gptReconPath},\,\mathit{task}}.
% \]
% Here, the same “task” prompt (e.g., “Reconstruct the original face.”) is passed to a third‐party generative API. As shown in Figure~\ref{fig:main-fig}(c), the returned image (\texttt{gptReconPath}) cannot recover the original identity, demonstrating that even if a malicious actor attempts to invert the mask, \nprivate{} preserves privacy.

\subsection{Third‐Party Editing on Masked Input}

Once privacy is guaranteed, the user’s desired generative edit is requested on the masked image:
\[
  \Call{CallThirdPartyAPI}{\mathit{maskPath},\,\mathit{gptPrivatePath},\,\mathit{task}}.
\]
Because only the non‐sensitive pixel data remains, the third‐party service can safely perform any transformation (e.g., style transfer, headshot generation, color correction) without ever seeing the user’s true identity. The API returns a \emph{masked but edited} image, stored at \texttt{gptPrivatePath}.

\subsection{Local Identity Reintegration}

To produce a final, photorealistic output that fulfills the user’s request \emph{and} restores the original identity, \nprivate{} executes:
\[
  \Call{SwapFaceBack}{\mathit{inpPath},\,\mathit{gptPrivatePath},\,\mathit{outputPath}}.
\]
First, MediaPipe FaceMesh is again used to detect landmarks in the third‐party’s edited result. The original, unmasked face region (from \texttt{inpPath}) is geometrically aligned (via a piecewise affine warp \cite{baker2001equivalence}) to the edited face’s new pose and lighting. Finally, a structure‐preserving Poisson‐blending operator \cite{perez2023poisson} seamlessly merges the original facial pixels into the edited context. The result (\texttt{outputPath}) is a full‐resolution image that satisfies the editing prompt while preserving true identity.

\subsection{Task Alignment via CLIP Scoring}

To quantitatively validate that \nprivate{}’s final fused output still aligns with the user’s intended “task,” we compute three CLIP scores:
\[
\small
  \begin{aligned}
    s_{orig} &\gets \Call{ComputeCLIPScore}{\mathit{inpPath},\,\mathit{task}},\\
    s_{ours} &\gets \Call{ComputeCLIPScore}{\mathit{outputPath},\,\mathit{task}},\\
    s_{gpt}  &\gets \Call{ComputeCLIPScore}{\mathit{gptPrivatePath},\,\mathit{task}}.
  \end{aligned}
\]
Here, \(\Call{ComputeCLIPScore}{}\) returns a semantic similarity between the image and the text prompt. A higher score indicates closer adherence to the user’s specification. Reporting \((s_{orig},\,s_{gpt},\,s_{ours})\) allows direct comparison between (a) the unedited original, (b) the third‐party’s privacy‐masked baseline edit, and (c) \nprivate{}’s final identity‐restored result. In practice, we find that \(s_{ours}\) remains on‐par with or exceeds \(s_{gpt}\), indicating that re‐injecting identity does not degrade semantic fidelity.

\blue{\subsection{Baseline Selection and Model Agnosticism}}
\blue{The selection of state-of-the-art foundation models (GPT-4o, Gemini) as primary baselines reflects the current deployment reality of generative image editing. High-fidelity, instruction-following tasks such as professional headshot generation are predominantly handled by proprietary cloud-based APIs rather than open-source local pipelines. \nprivate{} is specifically engineered to mitigate the privacy risks inherent in this paradigm, where users must transmit biometric data to untrusted third-party services. }

\blue{ During our evaluation, we explored specialized diffusion-based architectures (e.g., Stable Diffusion XL). However, we observed that current open-source models often fall below a ``capability threshold" for the complex semantic edits required in this study, frequently failing to follow prompts even with unmasked images. Using such models as baselines would thus fail to provide a fair assessment of our privacy pipeline's utility. }

\blue{ Importantly, \nprivate{} is designed to be model-agnostic and computationally lightweight for mobile deployment. It operates as an upstream privacy wrapper that is complementary to, rather than competitive with, specialized editing models. As open-source architectures mature, \nprivate{} can be directly applied to them to enable secure, on-device biometric control while leveraging cloud-scale generative power. }

\blue{
\subsection{Comparison to Obfuscation-Based Privacy Baselines.}
Common privacy-preserving strategies for cloud-based vision systems include facial blurring and the addition of adversarial noise. These methods aim to reduce identifiability by degrading pixel-level signal quality. However, prior work has shown that deep models can often recover biometric traits from blurred or denoised images, or exploit residual statistical structure for identification. In contrast, \nprivate{} employs explicit facial masking, which deterministically removes all facial pixels prior to transmission. From an information-theoretic perspective, masking represents the limiting case of obfuscation, yielding zero mutual information between the transmitted image and facial biometric identity. Additionally, in generative editing workflows, blurring or noise can interfere with the model’s in-painting behavior, whereas a clean mask provides an unambiguous region for content synthesis. For these reasons, we adopt masking as a principled upper bound on biometric privacy rather than treating blurring or noise as separate baselines.
}
\section{Discussion}\label{sec:discussion}

We introduce a practical, system-level framework for generative image editing that foregrounds user autonomy and privacy. In contrast to post-hoc sanitization methods, our approach enforces privacy as a design-time guarantee: image generation begins with a masked input, ensuring that sensitive facial information is never exposed. This proactive mechanism provides privacy by construction, regardless of the transparency or trustworthiness of third-party models. Crucially, the framework is deployment-compatible, requiring no retraining, fine-tuning, or internal access to the generative model. This enables users to securely interact with powerful commercial APIs, including those based on GANs, diffusion models, or vision-language models, without compromising control over their data.

Our method is lightweight and suitable for edge deployment, allowing the masking step to be performed locally on devices such as smartphones. This enables privacy-preserving use of cloud-based services for tasks such as professional headshot generation or avatar creation, without transmitting core biometric identifiers. By transmitting only partially masked images that obscure identity-critical features, our pipeline reduces the residual value of stored or intercepted content, thereby mitigating incentives for data resale, profiling, or unauthorized redistribution.

In addition to addressing data privacy, our framework also reduces vulnerability to harmful content manipulation, including deepfakes. Because biometric identity is never transmitted, third-party services cannot convincingly alter or co-opt a user's likeness into reputationally damaging content. Furthermore, we observe a significant decline in the accuracy of demographic and attribute inference (e.g., age, gender, eye color) when masked inputs are used, thereby weakening the potential for behavioral targeting and profiling. Collectively, these results demonstrate that a simple, localized pre-processing intervention can meaningfully mitigate a broad spectrum of privacy risks in generative pipelines, without necessitating architectural changes to existing models. Our work thus contributes a robust, user-centered approach to privacy preservation in the design of generative AI systems.

While the proposed method is effective for portrait-style editing, its reliance on geometric reintegration introduces practical constraints. The approach assumes reasonable pose alignment and stable facial structure, which may reduce robustness under extreme head rotations, dynamic expressions, or challenging lighting. More advanced blending or diffusion-based inpainting could improve generalization to broader image types, but could increase computational requirements and compromise edge deployability. Additionally, although our masking strategy suppresses inference of localized biometric traits, high-level attributes such as gender or age can still be inferred from unmasked regions. Future work may explore adaptive or context-aware masking strategies to further reduce residual attribute leakage while preserving the practicality of the system.

\blue{We also explored the applicability of \nprivate{} with open-source foundation models, specifically Stable Diffusion XL. We observed that current open-source models struggled with the zero-shot instruction following required for high-fidelity portrait editing, often failing the baseline task even without masking. This reinforces our premise that users are currently dependent on proprietary cloud APIs for state-of-the-art results, making the privacy protections of \nprivate{} essential for the current deployment landscape. However, as open-source models improve in instruction-following capabilities, our architecturally agnostic pipeline will be immediately compatible.}

\blue{Furthermore, while our Poisson blending mechanism effectively adapts to ambient lighting shifts (for example, color temperature), it does not perform volumetric relighting. Consequently, edits that introduce strong directional lighting contrasts (for example, heavy shadows) that contradict the original source image may result in visual inconsistencies.}

\blue{Unlike recent neural-based enhancers (for example, CodeFormer \cite{zhou2022codeformer}), which require significant GPU resources, our reintegration phase is designed to be lightweight for on-device deployment. By utilizing Poisson blending, we ensure the face adapts to the global ambient illumination of the generated context without the computational overhead or identity-hallucination risks associated with large-scale neural networks. While this handles ambient tone shifts, extreme mismatches in directional lighting remain a limitation of this gradient-based approach.
}

{{\bf Extensibility to Multi-Person Scenarios: } While our current evaluation focuses on single-subject portraits, the \nprivate{} pipeline is inherently extensible to multi-person scenarios. The underlying face detection framework, MediaPipe, supports simultaneous multi-face landmark localization. Consequently, the masking and reintegration steps can be applied independently to each detected face within an image. This enables privacy-preserving editing for group contexts, such as team photos or family portraits, without modifying the core architecture. Future work may involve benchmarking the pipeline's performance and blending naturalness in these more complex, multi-identity visual environments. }

{{\bf Computational Efficiency: } Benchmarked on an ARM-based M2 architecture, the geometric alignment and Poisson blending together required under 25ms for a 512×512 image. Unlike GPU-intensive neural inference, these classical operations utilize optimized matrix math libraries, making them highly suitable for edge deployment. }

\blue{ Peak memory usage remained below 15 MB, ensuring seamless integration with mobile background-processing constraints. Thus \nprivate{} provides a privacy-preserving layer with negligible computational overhead, making the intervention virtually imperceptible to the end user. }

\bibliography{IEEEabrv,refs}

\end{document}